\documentclass[12pt]{article}

\newcommand{\be}{\begin{equation}}
\newcommand{\ee}{\end{equation}}
\newcommand{\bea}{\begin{eqnarray}}
\newcommand{\eea}{\end{eqnarray}}
\def\abstract#1{\begin{center}{\large ABSTRACT}\end{center} \par #1}
\def\title#1{\begin{center}{\Large\bf {#1}}\end{center}}
\def\author#1{\begin{center}{\large #1}\end{center}}
\def\address#1{\begin{center}{\it #1}\end{center}}
 
\def\ft#1#2{{\textstyle{{\scriptstyle #1}\over {\scriptstyle #2}}}}

\renewcommand{\theequation}{\arabic{section}.\arabic{equation}}

\makeatletter
\@addtoreset{equation}{section}
\makeatother
\begin{document}%%%%%%%%%%%%%%%%%%%%%%%%%%%%%%%%%%%%
\begin{titlepage}
\hspace*{\fill}
\vbox{
\hbox{hep-th/0408030}
\hbox{KUNS-1932}
%  \hbox{\today}
\hbox{August 2004}
}
\vspace*{\fill}

\begin{center}
 \Large\bf 
Stringy effect of 
the holographic correspondence for D$p$-brane backgrounds 
\end{center}
\vskip 1cm
\author{
Masako {\sc Asano}%
\footnote{E-mail address:\ \ 
{\tt asano@gauge.scphys.kyoto-u.ac.jp}}}
\address{
Department of Physics, Kyoto University, \\
Kyoto 606-8502, Japan}

\vspace{\fill}
\abstract{
Based on the holographic conjecture 
for superstrings on D$p$-brane backgrounds 
and the dual $(p+1)$-dimensional gauge theory 
($0\le p\le 4$)
given in hep-th/0308024 and hep-th/0405203,
we continue the study of superstring amplitudes
including string higher modes ($n\ne 0$).
We give a prediction to the 
two-point functions of operators
with large R-charge $J$.
The effect of stringy modes do not appear as 
the form of anomalous dimensions except for $p=3$.
Instead, it gives 
non-trivial correction to the two-point functions 
for supergravity modes.
For $p=4$, 
the scalar two-point functions for any $n$ behave like free
fields of the effective dimension $d_{\rm eff}=6$
in the infra-red limit.
%which suggests the non-trivial fixed points.
}

\vspace*{\fill}

%       Keywords: 

\end{titlepage}
%%%%%%%%%%%%%%%%%%%%%%%%%%%%%%%%%%%%%%%%%%%%%%
\section{Introduction}
The BMN conjecture~\cite{Berenstein:2002jq}  
gives an important step toward the realization of 
the AdS/CFT correspondence~\cite{Maldacena:1997re} 
including stringy effect.
Originally,  AdS/CFT correspondence states that there is a 
holographic relation between 
bulk supergravity theory 
in the  AdS$_5 \times S^5 $ geometry 
and the boundary ${\cal N} = 4$ U($N$)
super Yang-Mills theory.
The relation between 
bulk field $\phi(x_\mu,z)$   
and the corresponding boundary operator ${\cal O}(x_\mu)$
is given explicitly by the GKP/Witten relation 
\cite{Gubser:1998bc, Witten:1998qj}.
On the other hand, BMN conjecture, which is a relation between 
the superstring amplitudes in the 
plane-wave limit of  AdS$_5 \times S^5 $ geometry and 
the ${\cal N} = 4$ SYM theory, does not seem to have such 
a holographic interpretation in its original form.
In ref.\cite{Dobashi:2002ar},
an interpretation of the 
BMN conjecture as a holographic relation of GKP/Witten type
is presented and it is further studied to 
construct a consistent string field theory~\cite{Dobashi:2004nm}.

Other than the conventional AdS$_5$/CFT$_4$ correspondence,
we can expect that such 
string/gravity correspondence can be applied 
to non-conformal case, such as the relation between 
D$p$-brane background
and the $(p+1)$-dimensional U($N$) SYM theory 
with 16 supercharges~\cite{Itzhaki:1998dd}.
For $p=0$, holographic correspondence between 
supergravity theory
and the $1$-dimensional 
SYM was studied~\cite{Sekino:1999av,Sekino:2000mg}
and the correspondence between supergravity states and SYM operators
is given based on the 
generalized conformal symmetry~\cite{Jevicki:1998yr}.

In our previous papers with Y.~Sekino and 
T.~Yoneya~ref.\cite{Asano:2003xp,Asano:2004vj}, 
extending the idea given in \cite{Dobashi:2002ar},
we investigated the closed superstring action 
around the null geodesic of D$p$-brane backgrounds 
and calculated the `diagonal' form of the boundary-to-boundary
$S$-matrix operator.
Our claim is that such $S$-matrix operator gives the 
two-point functions of the boundary gauge theory, 
which leads to the result consistent
with the BMN conjecture.
We mainly considered supergravity modes and 
obtained the two-point functions of certain operators ${\cal O}$
with large R-charge $J$
for boundary $(p+1)$-dimensional 
gauge theory and found that 
the result is consistent with the field theory analysis 
using the supergravity theory:
The two-point functions for supergravity modes 
show the power-law behavior and 
the contribution from the zero-point energies, 
which remain non-zero for $p\ne 3$, 
%because of the absence of world-sheet supersymmetry,
precisely agree with the supergravity results 
that are relevant for any $J$. 

Our aim of the present paper is to  
study the effect of 
string higher modes on the $S$-matrix and to
investigate the properties of two-point functions
of the boundary gauge theory 
predicted by the $S$-matrix.
We analyze the effect of $|n|\ne 0$ modes perturbatively 
around $|n|\to 0 $ or  $|n|\to \infty$.
%for $p\ne 3$. 
Then the UV or IR behavior of the two-point functions of 
dual gauge theory can be extracted. 
In particular, for $p=4$,  
the scalar two-point functions for any $n$ behave as for free
fields in the effective dimension $d_{\rm eff}=6$
in the infra-red limit,
which suggests the existence of non-trivial fixed points.

This paper is organized as follows.
In the next section, we 
review the superstring dynamics around the  
(tunneling) null geodesic for 
near horizon limit of D$p$-brane geometry.
Then we set up our problem to analyze, {\it i.e.,}
we explicitly write down the quadratic Hamiltonian $H=H_{\rm b}+H_{\rm f}$
representing the superstring fluctuations around the geodesic.
In section~3, we perform the quantization of 
the system and give the diagonalized form of $S$-matrix
operator.
In section~4, after reviewing the procedure that
gives the two-point functions 
$\langle \bar{\cal O}(x_f){\cal O}(x_i) \rangle$
in terms of the $S$-matrix
and the result for supergravity modes $n=0$, 
we analyze the effect of string higher modes $n\ne 0$
on the two-point functions. 
In section~4.1, we give the perturbative analysis of the $S$-matrix 
with respect to small $ |n|L^2/J $ and in section~4.2, we give the result for
$|n|L^2/J\to \infty$ limit.
Finally in section~4.3, we give the interpretation of the above result
as the correlation functions of the $(p+1)$-dimensional 
gauge theory.
In section 5, we give some concluding remarks.
In Appendix A, we explain the diagonalization procedure 
for the fermionic part of the $S$-matrix operator.

%%%%%%%%%%%%%%%%%%%%%%%%%%%%%%%%%%%%%%%%%%%%%%
\section{Superstring dynamics around the null geodesic in D$p$-brane backgrounds}
%%%%%%%%%%%%

D$p$-brane background for type II supergravity in the near-horizon limit 
is represented by the metric, 
Ramond-Ramond $(p\!+\!2)$-form field strength and the dilaton as
\begin{eqnarray}
ds^{2}&=&L^{2} \left[H^{-1/2}(-dt^{2}+d x_{a}^{2})+H^{1/2}
(dr^{2}+r^{2}(d\theta^2+ \cos^2\theta d\psi^2 + \sin^2\theta  
d\Omega_{6-p}^{2}))\right],
\nonumber\\
F_{p+2}& =& L^{p+1}\partial_{r} H^{-1}dt\wedge dx_{1}\wedge \cdots
\wedge 
dx_{p}\wedge dr,
\label{eq:nhDp}
\\
%e^{\phi}&=& g_{s}e^{\tilde\phi},\quad e^{\tilde\phi}=H^{\frac{3-p}{4}},
&& \quad  e^{\phi}= g_{s}H^{\frac{3-p}{4}},
\quad H = {1\over r^{7-p}} 
\nonumber
\end{eqnarray}
where $a=1,\ldots, p$, $L=q_{p}^{1/(7-p)}$ and 
$q_{p}=\tilde{c}_{p}g_{s}N\ell_{s}^{7-p}$ with 
$\tilde{c}_{p}=2^{6-p}\pi^{(5-p)/2}\Gamma{(7-p)/2}$.
We take $N \to \infty$ with fixed large $g_s N$
so that the coupling and curvature are small 
in the near-horizon region.
The coordinates we use are defined to be dimensionless
by rescaling $(t,x_{a},r)|_{\rm orig.} \to L(t,x_{a},r)$ from the original representation. 
Note that this background (or the supersrting action $S=S_{\rm b}+S_{\rm f}$ defined on 
the background) is invariant under the generalized scaling transformation
\begin{equation}
L\to \lambda^{\frac{3-p}{7-p}} L,
\quad
(t, x_a)\to \lambda^{-\frac{2(5-p)}{7-p}} (t, x_a)
,\quad
r \to \lambda^{\frac{4}{7-p}} r  .
\label{eq:scaletr}
\end{equation}
In the original coordinates, 
this transformation is represented as~\cite{Jevicki:1998yr,Asano:2003xp}
$g_s\to \lambda^{3-p} g_s$, 
$(t,x_a)|_{\rm orig.}\to \lambda^{-1} (t, x_a)|_{\rm orig.} $ and
$r_{\rm orig.} \to \lambda r_{\rm orig.}$.

We consider the Green-Schwarz superstring action in the above background
after performing `double Wick rotation' \cite{Dobashi:2002ar,Asano:2003xp}
for time and angle as $t\to -it$ $\psi\to -i\psi$.  
The bosonic part of the action is 
\begin{equation}
 S_{\rm b}={1\over 4\pi\alpha'}\int d\tau \int_{0}^{2\pi\tilde\alpha} d\sigma
\sqrt{h} h^{\alpha\beta}\partial_{\alpha}x^{\mu}
\partial_{\beta}x^{\nu} \tilde{g}_{\mu\nu}
\label{eq:Sb}
\end{equation}
where $\alpha,\beta$ denote the world-sheet coordinates
$\tau,\sigma$ with signature $(+,+)$
and $\tilde{\alpha}$ denotes the world-sheet length scale which 
will be fixed later. 
The metric $\tilde{g}_{\mu\nu}$ is given by (\ref{eq:nhDp})
after the double Wick rotation.
We set $\alpha'=1$ hereafter.

We consider a point-like classical solution of the above action
with $x_a=\theta=0$, which has
conserved energy and angular momentum along $\psi$:
\begin{equation}
E = L^2 r^{7-p\over 2} \dot{t} \sqrt{h} h^{\tau\tau} \tilde{\alpha}
,\quad
J = L^2 r^{-{3-p\over 2}} \dot{\psi} \sqrt{h} h^{\tau\tau} \tilde{\alpha}
.
\end{equation}
By choosing the gauge for the world-sheet metric and $\tilde{\alpha}$ as 
\begin{equation}
\sqrt{h} h^{\tau\tau} =
\left(\frac{\cosh\!\tau}{\ell} \right)^{(3-p)/(5-p)},
\quad 
 \tilde\alpha={5-p\over 2}{J\over L^{2}}
%\bar{r}(\tau)^{3-p \over 2},
\label{eq:hgaugeun}
\end{equation} 
the solution is written as
\begin{equation}
 z={\tilde\ell\over \cosh \tau},\quad 
t=\tilde\ell \tanh \tau,\quad \psi={2\over 5-p}\tau
\label{eq:trajP}
\end{equation}
where $z={2\over 5-p}r^{-(5-p)/2}$, $\ell\equiv J/E$
and $\tilde{\ell}={2\over 5-p } \ell $.
For $0 \le p \le 4$,
this solution represents the null geodesic connecting two points 
$t=t_i$ and $t=t_f$
on the $(p+1)$-dimensional boundary%
\footnote{By $z\sim 0$, we mean $z=1/\Lambda$ with 
$\Lambda\tilde{\ell}\to \infty$.}
 $z\sim 0$ of $(p+2)$-dimensional 
space $({t}, z,x_{a})$ which is conformal to AdS$_{p+2}$.
The separation between the two end-points at $z(=1/\Lambda)\sim 0$
is $|t_f-t_i| \sim 2\tilde{\ell}$.
We sometimes call this geodesic  `tunneling null geodesic.'
In our previous two papers \cite{Asano:2003xp,Asano:2004vj}
with Y.~Sekino and T.~Yoneya,
we claimed that the two-point functions 
$\langle {\cal O}(t_i) {\cal O}(t_f) \rangle$
of BMN type operators ${\cal O}$ can be obtained 
from the investigation of various modes of the 
string fluctuations around the geodesic.  

To illustrate this discussion explicitly, we
first present the quadratic action for fluctuations around the geodesic.
It is given by expanding the Green-Schwarz action around the 
classical solution $\Xi=\bar{\Xi}(\tau)$ as
$$\Xi(\tau,\sigma) = \bar{\Xi}(\tau) + {\Xi^{(1)}(\tau,\sigma)  \over L} 
+ {\Xi^{(2)}(\tau,\sigma)  \over L^2}+\cdots$$ 
where $\Xi$ is each bosonic or fermionic field appearing in the action. 
%and taking the leading order of $1/L$.
By taking the gauge (\ref{eq:hgaugeun}), the resulting bosonic action 
up to quadratic order is
\begin{eqnarray}
S_{\rm b}^{(2)}={1\over 4 \pi}
\int d\tau \int_0^{2\pi \tilde{\alpha}  } \!\!\! d\sigma\,
\Big[
\left(\ft{2}{5-p}\right)^2 (\dot{x}_i^2+\bar{r}(\tau)^{p-3} x_i'^{\,2})
+\dot{y}_l^2+\bar{r}(\tau)^{p-3} y_l'^{\,2}
\nonumber\\
\hspace*{5cm}
+\left(\ft{2}{5-p}\right)^2
(x_i^2 +y_l^2)
\Big] \quad +\, {\cal O}(L^{-1})
\label{eq:actionbos2}
\end{eqnarray}
where 
 $8(=[p\!+\!1]+[7\!-\!p])$ fields $x_{i}$
$(i=1,\ldots,p+1)$ and $y_{l}$ $(l=1,\ldots, 7-p)$ 
are dynamical fields remaining after gauge fixing 
among 10 original bosonic fields within $\Xi^{(1)}$. 
The $\sigma$-dependence can be Fourier transformed by
taking 
\begin{equation}
X(\tau,\sigma)={1\over \sqrt{\tilde{\alpha}}}
\, \sum_{n\ge 0} \, \left[
\cos\Big( {n\over \tilde{\alpha}}\sigma \Big) X_n(\tau)
 + \sin\Big( {n\over \tilde{\alpha}}\sigma \Big) X_{-n} (\tau)
\right]
\end{equation}
where $X$ denotes $x_i$ or $y_l$.
The equation of motion for each mode is
\begin{equation}
\ddot{X}_n  -\left[
\bar{r}(\tau)^{p-3} \Big( {n\over \tilde{\alpha}} \Big)^2
+m^2 
\right] X_n = 0
\label{eq:EOMbos}
\end{equation}
where 
\begin{equation}
\bar{r}(\tau)=
\left(\frac{\cosh\!\tau}{\ell} \right)^{2/(5-p)},
\end{equation}
$m=1$ for $x_i$ and $m=\frac{2}{5-p}$ for $y_l$.
We sometimes denote 
$$
M_n(\tau)^2 \equiv \bar{r}(\tau)^{p-3} \Big( {n\over \tilde{\alpha}} \Big)^2 +m^2 .
$$
Note that $M_n(\tau)^2 = M_n(-\tau)^2$.
Hamiltonian for each mode after some rescaling of field is given as
\begin{equation}
H_n^X=\frac{1}{2} \left(
P_n^2 +M_n(\tau)^2 X_n^2
\right)
\end{equation}
where $P_n (= -i \partial{\cal L}/\partial \dot{X}_n ) = -i \dot{X}_n$.
Thus the total Hamiltonian for bosonic fluctuations is
given by summing up all modes from $x_i$ and $y_l$ as
\begin{equation}
H_{\rm b}= \sum_{I=(i,l)}\sum_{n=-\infty}^{\infty} H_n^I.
\end{equation}

%fermion

For fermionic fluctuations, the quadratic action around the trajectory
is also obtained by
expanding the GS action~\cite{Cvetic:2002nh}.
The result for IIA or IIB is given as~\cite{Asano:2004vj}
\begin{eqnarray}
 S^{(2)}_{{\rm f},A}&=& {1 \over 2\pi} \int d\tau
\int^{2\pi\tilde{\alpha}}_{0} \! d\sigma \,
\Theta^{T}\Gamma_{0}
\Bigg[ \sqrt{2} \Gamma_{+} \partial_{\tau}\Theta -i m_{f(p)}
%{7-p \over 2(5-p)}
\Gamma_{+} \Gamma_{(p)} \Gamma_{+} \Theta
\nonumber\\
&& \hspace{7cm} -i \sqrt{2}\,
%\left( {\cosh\tau \over \ell} \right)^{-(3-p)/(5-p)}
\bar{r}(\tau)^{-{3-p \over 2}}
\Gamma_{11} \Gamma_{+}
\partial_{\sigma} \Theta\Bigg],
\label{eq:SfA2}
\end{eqnarray}
\begin{eqnarray}
 S^{(2)}_{{\rm f},B}&=& {1 \over 2\pi} \int d\tau
\int^{2\pi\tilde{\alpha}}_{0} \! d\sigma \,
(\Theta^I)^{T} \Gamma_{0}
\Bigg[ \sqrt{2} \Gamma_{+} \partial_{\tau}\Theta^I -i  m_{f(p)}
\Gamma_{+} \Gamma_{(p)}^{IJ} \Gamma_{+} \Theta^J
\nonumber\\
&& \hspace{7cm} +i \sqrt{2}\,
%\left( {\cosh\tau \over \ell} \right)^{-(3-p)/(5-p)}
\bar{r}(\tau)^{-{3-p \over 2}}
s_2^{IJ} \Gamma_{+}
\partial_{\sigma} \Theta^J  \Bigg]
\label{eq:SfB2}
\end{eqnarray}
where $\Theta$ for IIA is 32-component Majorana spinor and 
$\Theta^I$ ($I=1,2$) for IIB are two Majorana-Weyl spinors.
Matrices $s_{k}^{IJ}$ ($k=0,1,2$) are given by Pauli matrices as
$s_{0}=-i\sigma_{2}$, $s_{1}=\sigma_{1}$ and $s_2=\sigma_3$.
We choose the representation of Gamma matrices as 
\begin{eqnarray}
&&\Gamma_{0} = \pmatrix{0 & 1\cr -1& 0},\quad 
\Gamma_{+} = \pmatrix{0 & \sqrt{2}\cr 0& 0},\quad 
\Gamma_{\hat{x}_i} = \pmatrix{\gamma_i & 0\cr 0& -\gamma_i} ,
\nonumber\\
&&
\Gamma_{\hat{y}_l} = \pmatrix{\gamma_{p+1+l} & 0\cr 0& -\gamma_{p+1+l}} ,\quad 
\Gamma_{11}=\pmatrix{\gamma_9&0\cr 0& -\gamma_9}\,
%&&
%\Gamma_+= \pmatrix{0& \sqrt2\cr 0& 0}\,,\;
%\Gamma_-= \pmatrix{0& 0\cr \sqrt2 & 0}\;
\label{eq:repGamma}
\end{eqnarray}
and decompose $\Theta$ as 
$$
\Theta={1\over \sqrt{2}}\pmatrix{\hat\theta \cr \theta}\, .
$$
We see that only $\theta$ components appear in the action (\ref{eq:SfA2}) or   
(\ref{eq:SfB2}). 
If we further decompose $\theta$ as
\begin{eqnarray}
&& \theta=\theta_+ + \theta_- ,\quad \gamma_9 \theta_{\pm} = \pm \theta_{\pm} 
 \hspace*{2.45cm}  \mbox{for IIA},
\\
&& (\theta^1, \theta^2) = (\theta_-, \theta_+) \quad \mbox{or}\quad 
 s_2 \theta_{\pm} = \mp \theta_{\pm} 
 \qquad  \mbox{for IIB},
\end{eqnarray}
the action for IIA and IIB is represented 
in a unified form  
\begin{eqnarray}
S^{(2)}_{\rm f} &=&  -{1\over 2\pi}\int d\tau 
\int^{2\pi \tilde{\alpha} }_{0}
\! d\sigma \, \Big[
\theta_+^{T} \partial_{\tau} \theta_+ 
+ \theta_-^{T} \partial_{\tau} \theta_-
\nonumber\\
&& \hspace*{2cm} - 2 i m_{f(p)} \theta_+^{T} \gamma_{(p)} \theta_-
-i \bar{r}(\tau)^{-{3-p \over 2}}
( \theta_+^T \partial_{\sigma} \theta_+ -  \theta_-^T \partial_{\sigma} \theta_- )
\big].
\end{eqnarray}
Here 
\begin{equation}
m_{f(p)}={7-p\over 2(5-p)}  
\end{equation}
and 
\begin{equation}
 \gamma_{(p=0)}= \gamma_{9}\gamma_{1},\quad
 \gamma_{(p=2)}= \gamma_{123},\quad
 \gamma_{(p=4)}= -\gamma_{9}\gamma_{12345},
\end{equation}
\begin{equation}
 \gamma_{(p=1)}= s_1 \gamma_{12},\quad
 \gamma_{(p=3)}= - s_0 \gamma_{1234}.
\end{equation}
The equation of motion for each 
Fourier mode $\theta_{\pm,n}(\tau)$ given by 
\begin{equation}
\theta_{\pm}(\tau,\sigma)= {1\over \sqrt{\tilde{\alpha}}}\,
\sum_{n=-\infty}^{\infty}
\theta_{\pm,n}(\tau) e^{i {n\over \tilde{\alpha}} \sigma}
\end{equation}
are 
\begin{equation}
\partial_{\tau} \theta_{\pm,n} 
\pm {n\over \tilde{\alpha}} \, \bar{r}(\tau)^{-{3-p \over 2}} \theta_{\pm,n} 
-  i m_{f(p)} \gamma_{(p)} \theta_{\mp,n}=0.
\label{eq:EOMfer}
\end{equation}
Hamiltonian is given as
\begin{equation}
H_{\rm f} =   \sum_{n=-\infty}^{\infty} \bigg[
- 2 i m_{f(p)} \theta_{+,-n}^{T} \gamma_{(p)} \theta_{-,n}
+ {n\over \tilde{\alpha}} \, \bar{r}(\tau)^{-{3-p \over 2}}
( \theta_{+,-n}^T \theta_{+,n} -  \theta_{-,-n}^T  \theta_{-,n} )
\bigg].
\end{equation}
Thus we have found complete Hamiltonian $H=H_{\rm b}+H_{\rm f}$ for all the 
quadratic fluctuations around the classical trajectory.

Finally, note that 
the final form of the action $S = S^{(2)}_{\rm b} + S^{(2)}_{\rm f}$
coincides with the one obtained for the fluctuations around the 
real null geodesic in D$p$-brane geometry without double Wick rotation.

%%%%%%%%%%%%%%%%%%%%%%%%%%%%%%%%%%%%%%%%%%%%%%
\section{Quantization and the Diagonalization of $S$-matrix}
%%%%%%%%%%%%%%%
\subsection{$S$-matrix from $\tau=-T$ to $\tau=T$: definition}

Now we quantize the system $H=H_{\rm b}+H_{\rm f}$ 
and analyze the boundary-to-boundary
$S$-matrix along the tunneling null geodesic, which is 
interpreted as two-point functions of the boundary gauge theory.

For bosonic sector, we write general solutions of (\ref{eq:EOMbos}) as
\begin{equation}
X_n(\tau) = f_n^{(+)}(\tau) a_n + f_n^{(-)}(\tau) a_n^\dagger
\label{eq:Xsolbos}
\end{equation}
with constant operators $a_n$ and $a_n^\dagger$.
We choose $f_n^{(\pm)}$ to satisfy 
the time reflection symmetry~\cite{Asano:2003xp} 
$f_n^{(+)}(\tau) = f_n^{(-)}(-\tau)$ 
which is 
alternative to the reality condition in the real-time formulation. 
We also set the `boundary condition' at $\tau \to \infty$ as 
$f^{(+)}_n \to 0$ (or at least $f^{(+)}_n/f^{(-)}_n \to 0$) 
with divergent $f_n^{(-)}$.
Furthermore, we impose 
the normalization condition 
\begin{equation}
f_n^{(+)}  {d f_n^{(-)}  \over d\tau} 
- f_n^{(-)}  {d f_n^{(+)} \over d\tau} =1 .
\label{eq:normcondbos}
\end{equation}
Then, the canonical commutation relation $[X_n,P_{n'}]=i\delta_{n, n'}$ 
becomes equivalent to 
$[a_n,a_{n'}^\dagger]=\delta_{n,n'}$.
The Hamiltonian is written by $(a_n,a_n^\dagger)$ as 
\begin{eqnarray}
H_{\rm b} &=& \sum_{I=(i,l)}\sum_{n=-\infty}^{\infty} \Bigg\{
 \frac{1}{2} \bigg[ - ( \dot{f}_n^{I,(+)})^2 
+ (M_n^I)^2 (f_n^{I,(+)})^2 \bigg] (a_n^I)^2
\nonumber\\ 
&& \hspace*{2cm} 
+ \frac{1}{2} \bigg[ -\dot{f}_n^{I,(+)} \dot{f}_n^{I,(-)} 
+ (M_n^I)^2 f_n^{I,(+)} f_n^{I,(-)} \bigg] 
\bigg( a_n^{I\dagger} a_n^I + \frac{1}{2}\bigg) 
\nonumber\\ 
&& \hspace*{2cm}
+  \frac{1}{2} 
\bigg[ - (\dot{f}_n^{I,(-)})^2 
+ (M_n^I)^2 (f_n^{I,(-)})^2 \bigg] (a_n^{I\dagger})^2
\Bigg\}.
\end{eqnarray}

%%%
For fermionic sector, general solutions for
(\ref{eq:EOMfer}) are expressed by spinor operators $d_n^\alpha$ and 
$d_n^{\alpha\dagger}$ $(\alpha=1,\cdots,8)$ as
\begin{equation}
\left(\theta_{-,n} (\tau) , \tilde{\theta}_{+,n}(\tau) \right)
= \left( \phi_n^{(+)}(\tau) \,d_n+  \phi_n^{(-)}(\tau)  \,d_n^\dagger \;,\;  
\psi_n^{(+)}(\tau) \, d_n+  \psi_n^{(-)}(\tau)\,  d_n^\dagger  \right)
\label{eq:fermigensol}
\end{equation}
where $ \tilde{\theta}_{+,n}=  i \gamma_{(p)} \theta_{+,n}$. 
We choose 
\begin{equation}
\phi_n^{(+)} \psi_n^{(-)} - \phi_n^{(-)} \psi_n^{(+)} = {1\over 2}
\label{eq:solcond1}
\end{equation}
and 
\begin{equation}
\phi_n^{(+)} =- \psi_{-n}^{(+)},\quad  \phi_n^{(-)} = \psi_{-n}^{(-)} .
\label{eq:solcond2}
\end{equation}
Also, we can set  $ \phi_{-n}^{(+)} (-\tau) = \phi_n^{(-)}(\tau)$ 
from the time reflection symmetry $\theta_{\pm,n}(\tau)^\dagger = \theta_{\pm, -n}(-\tau)$
\cite{Asano:2004vj}. 
Furthermore, we set the boundary condition at $\tau \to \infty$
as $\phi_{n}^{(+)} \to 0$ or $\phi_{n}^{(+)}/\phi_{n}^{(-)} \to 0$
with $|\phi_n^{(-)}|\to \infty$.
Then the canonical anti-commutation relations for $\theta_{\pm,n}$ 
\begin{equation}
\{\theta_{s,n}^\alpha, \theta_{s',n'}^{\alpha'} \} = 
{1\over 2} \delta_{s,s'} \delta_{n,-n'} \delta_{\alpha,\alpha'} 
\label{eq:anticomm}
\end{equation}
lead to 
\begin{equation}
\{d_n^\alpha, d_{n'}^{\alpha'\dagger} \} = \delta_{n,-n'} \delta_{\alpha,\alpha'}.
\end{equation}
The Hamiltonian is written as
\begin{eqnarray}
H_{\rm f} &=& \sum_{n=-\infty}^{\infty} \sum_{\alpha=1}^8 
 \Bigg\{
- 2 \bigg[ m_{f(p)} \Big( \phi_{n}^{(+)} \Big)^2
+{n\over \tilde{\alpha}} \bar{r}^{{p-3 \over 2}} 
\phi_{n}^{(+)} \phi_{-n}^{(+)} 
 \bigg] d_{-n}^{\alpha} d_n^\alpha  
\nonumber\\
& &  + 2 \bigg[ m_{f(p)} \Big( \phi_{n}^{(+)} \phi_{n}^{(-)} + \phi_{-n}^{(+)} \phi_{-n}^{(-)} \Big)
+{n\over \tilde{\alpha}} \bar{r}^{{p-3\over 2}} 
\Big(\phi_{-n}^{(+)} \phi_{n}^{(-)} - \phi_{n}^{(+)} \phi_{-n}^{(-)} \Big)
 \bigg] \bigg( d_{-n}^{\alpha\dagger} d_n^\alpha-{1\over 2}  \bigg) 
\nonumber\\
&& + 2 \bigg[ m_{f(p)} \Big( \phi_{n}^{(-)} \Big)^2
-{n\over \tilde{\alpha}} \bar{r}^{{p-3 \over 2}} 
\phi_{n}^{(-)} \phi_{-n}^{(-)} 
 \bigg] d_{-n}^{\alpha\dagger} d_n^{\alpha\dagger}  
\Bigg\}.
\end{eqnarray}

We define the (Euclidean) $S$-matrix from $\tau=-T$ to $\tau=T$ as
the integration of the anti-time ordered product
\begin{equation}
 S(T)={\cal T}_{-} \exp\left[ -\int^{T}_{-T} d\tau H(\tau)\right]  
\label{eq:defSint}
\end{equation}
or 
\begin{equation}
\frac{dS(T)}{dT} =-H(-T) S(T) -S(T) H(T).
\label{eq:defSdiff}
\end{equation}
Note that $S(T)$ is hermitian $S^\dagger(T)=S(T)$ since
$H^\dagger(\tau)=H(-\tau)$. 
(We assume $f_n^{(\pm)}$ and $\phi_n^{(\pm)} $ are real.)  
We will interpret this $S$-matrix as the diagonalized  
two-point functions of BMN-type operators ${\cal O}$
of the boundary gauge theory. 
In general, this Hamiltonian $H=H_{\rm b}+H_{\rm f}$
is time-dependent $H=H(\tau)$ and $S(T)$ must be `diagonalized'
by performing time-dependent Bogoliubov transformation 
in order to extract the information of the diagonalized value 
of two-point functions for the dual gauge theory.

In the following, we review the diagonalization procedure 
for bosonic part of the $S$-matrix developed in ref.\cite{Asano:2003xp}
and then generalize the discussion to the fermionic part.

%%%%%%%%%%%%%%%
\subsection{General theory for time-dependent
Harmonic oscillators}
\paragraph{Bosonic part}
We consider each mode separately by 
decomposing the $S$-matrix as
\begin{equation}
S_{\rm b}(T) = \prod_{I=x_i,y_l}\prod_{n=-\infty}^{\infty}S_{I,n}(T)   
\label{eq:defSb}
\end{equation}
where 
\begin{equation}
S_{\rm b}^{I,n}(T)   =  \prod_{n,I} 
 {\cal T}_- \exp \left[ -\int^{T}_{-T} d\tau \, H_{\rm b}^{n,I}(\tau)\right] .
\end{equation}
As was discussed in ref.\cite{Asano:2003xp}, we can represent 
$S_{\rm b}(T)$ (or $S_{I,n}$) in 
two ways. 
One is normal ordered form which is naturally obtained from 
the definition of $S$-matrix (\ref{eq:defSint}) or (\ref{eq:defSdiff}) as
\begin{equation}
S_{I,n} (T) =
N_{I,n} (T)
 : \exp\left[{1\over 2}A_{I,n}(T) (a^{\dagger}_{I,n} )^2
+ B_{I,n} (T) a^{\dagger}_{I,n} a_{I,n}  
+ {1\over 2}C_{I,n}(T) a^2_{I,n} \right]: \, 
\end{equation}
with 
\begin{equation}
N_{I,n}^2=1+B_{I,n}=\frac{1}{2 f_{I,n}^{(-)} \dot{f}_{I,n}^{(-)} },
\quad
A_{I,n}=C_{I,n}=-\frac{1}{2}\left( \frac{f_{I,n}^{(+)}}{f_{I,n}^{(-)}} 
+ \frac{\dot{f}_{I,n}^{(+)}}{\dot{f}_{I,n}^{(-)}}    \right).
\end{equation}
Another is exponential form 
\begin{equation}
S_{I,n} (T) =
\tilde{N}_{I,n} (T)\exp\left[{1\over 2}\tilde{A}_{I,n} (T)
(a^{\dagger}_{I,n} )^2
+\tilde{B}_{I,n} (T)a^{\dagger}_{I,n} a_{I,n}  
+{1\over 2}\tilde{A}_{I,n} (T)a^2_{I,n} \right]
\label{eq:Sbexp}
\end{equation}
where $\tilde{A}_{I,n}$, $\tilde{B}_{I,n}$ and 
$\tilde{N}_{I,n}$
are determined by
\begin{equation}
\exp\left(
\begin{array}{cc}
-\tilde{B}_{I,n} & -\tilde{A}_{I,n}
\\ \tilde{A}_{I,n} & \tilde{B}_{I,n} 
\end{array}
\right)
=
\frac{1}{1+B_{I,n}}
\left(
\begin{array}{cc}
1 & -A_{I,n}
\\ 
A_{I,n} & (1+B_{I,n})^2 -A_{I,n}^2 
\end{array}
\right)
\end{equation}
and%
\footnote{We correct a minor mistake of eq.(3.42) in \cite{Asano:2003xp}.} 
\begin{equation}
\tilde{N}_{I,n}=N_{I,n} \exp(\tilde{B}_{I,n}/2) (1+B_{I,n})^{-1/2} .
\end{equation}
Once $S(T)$ is represented by exponential form,
it can be transformed to a `diagonalized' form by 
$T$-dependent Bogoliubov transformation
\begin{equation} 
\left(
\begin{array}{c}
a^\dagger_{I,n}  \\ a_{I,n}  
\end{array}
\right)
\to
\left(
\begin{array}{c}
b_{I,n}^\dagger(T) \\ b_{I,n} (T)
\end{array}
\right)
=
\left(
\begin{array}{cc}
G_{I,n}(T)  & F_{I,n}(T)
\\ E_{I,n}(T) & D_{I,n}(T) 
\end{array}
\right)
\left(
\begin{array}{c}
a^\dagger_{I,n} \\ a_{I,n} 
\end{array}
\right).
\end{equation}
We choose $D$, $E$, $F$ and $G$ ($DG-EF=1$) in order to satisfy 
\begin{eqnarray}
S_{I,n} (T) &=& \tilde{N}_{I,n} 
\exp\left(- {\tilde{B}_{I,n} \over 2}\right)
\exp
\left[
\frac{1}{2} \left(a^\dagger_{I,n} \; a_{I,n} \right) 
\left(
\begin{array}{cc}
\tilde{A}_{I,n} & \tilde{B}_{I,n}
\\ \tilde{B}_{I,n} & \tilde{A}_{I,n} 
\end{array}
\right)
\left(
\begin{array}{c}
a^\dagger_{I,n} \\ a_{I,n} 
\end{array}
\right)
\right]
\nonumber\\
&=&
\tilde{N}_{I,n}
\exp\left(- {\tilde{B}_{I,n} \over 2}\right)
\exp
\left[
-\frac{1}{2} \Big(b^\dagger_{I,n} \; b_{I,n}\Big) 
\left(
\begin{array}{cc}
0 & \Omega_{I,n}
\\ \Omega_{I,n} & 0 
\end{array}
\right)
\left(
\begin{array}{c}
b^\dagger_{I,n} \\ b_{I,n} 
\end{array}
\right)
\right]
\end{eqnarray}
where
\begin{equation}
\Omega_{I,n} = \sqrt{  \tilde{B}_{I,n}^2 -\tilde{A}_{I,n}^2  }\,
\end{equation}
or
\begin{equation}
\cosh \Omega_{I,n} = \frac{1}{2}
\left(
1+B_{I,n} +\frac{1-A_{I,n}^2}{1+B_{I,n}} 
\right).
\end{equation}
The final form of diagonal $S$-matrix is 
\begin{equation}
S_{\rm b} (T) = \prod_{I,n}\exp \left[-\Omega_{I,n} 
\left(b^\dagger_{I,n}(T) \, b_{I,n}(T)  + \ \frac{1}{2}\right) \right]
\end{equation}
where
\begin{equation}
[b_{I,n}(T),b_{I',n'}^\dagger(T)]=\delta_{I,I'}\delta_{n,n'}.
\end{equation}

Since we have chosen 
$f^{(+)}_{I,n}/f^{(-)}_{I,n} \to 0$
and $|f^{(-)}_{I,n}| \to \infty $ 
in the $T\to \infty$ limit, $A\sim 1+B \to 0$ 
and $(b^{\dagger} (T), b(T)) \to (a^\dagger,a)$.
Thus,
\begin{equation}
S_{\rm b} (T) \stackrel{T\to \infty}{\longrightarrow}
\prod_{I,n} \left(
2 f^{(-)}_{I,n}(T) \dot{f}^{(-)}_{I,n}(T) 
\right)^{-\left(a_{I,n}^\dagger a_{I,n} +\frac{1}{2}\right)}.
\end{equation}

%%%%%%%%%%%%%%%
\paragraph{Fermionic part}
Now we discuss the fermionic part of $S$-matrix 
\begin{equation}
S_{\rm f}(T) = \prod_{\alpha=1}^8 \prod_{n=-\infty}^{\infty}
{\cal T}_- \exp \left[
-\int^T_{-T} d\tau H_{\rm f}^{n}
\right]
\left(
\equiv \prod_{\alpha,n}S_{n}(T)
\right).
\end{equation}
As in the case of bosonic part, 
the first step is to write $S_{n}$ in the normal-ordered form.
We can write 
\begin{equation}
S_{n} (T) =
N^{\rm f}_{n} (T)
 : \exp\left[{1\over 2}A^{\rm f}_{n}(T) d^{\alpha \dagger}_{n} 
d^{\alpha \dagger}_{-n} 
+ B^{\rm f}_{n} (T) d^{\alpha \dagger}_{n} d^{\alpha}_{-n}  
+ {1\over 2} C^{\rm f}_{n}(T) d^{\alpha}_{n} d^{\alpha}_{-n}\right]: \, 
\label{eq:normalof}
\end{equation}
with 
\begin{equation}
A^{\rm f}_n =-A^{\rm f}_{-n}, \quad B^{\rm f}_n = B^{\rm f}_{-n}, \quad 
C^{\rm f}_n =-C^{\rm f}_{-n} , \quad N^{\rm f}_n = N^{\rm f}_{-n}. 
\label{eq:n-nrel}
\end{equation}
Note that $S_{n}$ and $S_{-n}$ 
do not commute with each other
and we have to treat them pairwise.
From the definition, $S$-matrix satisfies the relation
(\ref{eq:defSdiff})
and
\begin{equation}
\theta_{-,n} (T) =S(T)^{-1} \theta_{-,n} (-T) S(T), \quad
\tilde{\theta}_{+,n}  (T) =S(T)^{-1} \tilde{\theta}_{+,n} (-T) S(T).
\end{equation}
These relations determine $A_n^{\rm f}$, $B_n^{\rm f}$, $C_n^{\rm f}$ 
and $N_n^{\rm f}$ as 
\begin{equation}
1+B^{\rm f}_{n} = \frac{1}{\big(N^{\rm f}_{n} \big)^2}
 =\frac{1}{2}\,
\frac{1}
{ \big( \phi_{n}^{(-)} \big)^2
+
\big( \phi_{-n}^{(-)} \big)^2
 },
\quad
A^{\rm f}_{n}=C^{\rm f}_{n}=
 \frac{\phi_{-n}^{(+)} \phi_{-n}^{(-)}- \phi_{n}^{(+)} \phi_{n}^{(-)}  }
{\big( \phi_{n}^{(-)} \big)^2 + \big( \phi_{-n}^{(-)} \big)^2}    .
\end{equation}
Second step is to represent $S_{n}(T)$ in the exponential form 
\begin{equation}
S_{\rm f} (T) =
\left(\prod_{\alpha,n} \tilde{N}^{\rm f}_{n} (T)\right)
 \exp\left\{\sum_{\alpha,n}
\left[{1\over 2} \tilde{A}^{\rm f}_{n}(T) 
d^{\alpha\dagger}_{n} d^{\alpha\dagger}_{-n} 
+ \tilde{B}^{\rm f}_{n} (T) d^{\alpha\dagger}_{n} d^{\alpha}_{-n}  
+ {1\over 2} \tilde{A}^{\rm f}_{n}(T) d^{\alpha}_{n} d^{\alpha}_{-n} 
\right]\right\} \, 
\label{eq:expof}
\end{equation}
where $\tilde{A}_n^{\rm f}$, 
$\tilde{B}_n^{\rm f}$ and $\tilde{N}_n^{\rm f}$
also satisfy the relation corresponding to (\ref{eq:n-nrel}).
The relation between $(A_n^{\rm f}, B_n^{\rm f},N_n^{\rm f})$ and
 $(\tilde{A}_n^{\rm f}, \tilde{B}_n^{\rm f}, \tilde{N}_n^{\rm f})$ 
is determined explicitly in Appendix~A and the result is
\begin{eqnarray}
\tilde{N}^{\rm f}_n &=&
N^{\rm f}_n  \exp\left(-{\tilde{B}^{\rm f}_{n}\over 2}\right) \sqrt{1+B^{\rm f}_{n}} 
\\
\cosh \Omega^{\rm f}_{n} & = & \frac{1}{2}
\left(
1+B^{\rm f}_{n} +\frac{1+(A^{\rm f}_{n})^2}{1+B^{\rm f}_{n}} 
\right),
\\
\tilde{B}^{\rm f}_n \frac{\sinh \Omega^{\rm f}_{n} }{\Omega^{\rm f}_n }
&=& \frac{1}{2}
\left(
1+B^{\rm f}_{n} -\frac{1-(A^{\rm f}_{n})^2}{1+B^{\rm f}_{n}} 
\right),
\\
\tilde{A}^{\rm f}_n \frac{\sinh \Omega^{\rm f}_{n} }{\Omega^{\rm f}_n }
&=& \frac{A^{\rm f}_{n}}{1+B^{\rm f}_{n}} 
\end{eqnarray}
with 
\begin{equation}
%\tilde{A}^{\rm f}_{n} = \tilde{C}^{\rm f}_{n},\quad
\Omega^{\rm f}_{n} = \sqrt{(\tilde{A}^{\rm f}_n)^2 + (\tilde{B}^{\rm f}_n)^2}\, .
\end{equation}
Note that the above equations 
have a symmetry for $\Omega^{\rm f}_{n} \to -\Omega^{\rm f}_{n}$:
We fix the ambiguity by imposing
the condition
$\Omega^{\rm f}_{n} (T\to \infty)\rightarrow  \infty$. 
Finally, 
we can convert this to the diagonalized form  
\begin{equation}
S_{\rm f} (T) = \prod_{\alpha,n}\exp \left[-\Omega^{\rm f}_{n} (T)
\left(c^{\alpha\dagger}_{n}(T) \, c^{\alpha}_{-n}(T) 
 - \ \frac{1}{2}\right) \right]
\label{eq:diagSf}
\end{equation}
with $T$-dependent operators $c^\alpha_n$
and $c^{\alpha\dagger}_{n}$ satisfying
\begin{equation}
\{ c^\alpha_n(T) , c^{\alpha'\dagger}_{n'}(T)  \}
= \delta_{\alpha, \alpha'}\delta_{n,-n'}.
\label{eq:anticomc}
\end{equation}

For $T\to \infty$, 
$ \exp (\Omega^{\rm f}_{n}) \rightarrow  1/(1+ B_n^{\rm f}) $ since
we have chosen $ \phi^{(+)}_n / \phi^{(-)}_n \to 0 $ 
and
$|\phi^{(-)}_n| \to \infty$ in this limit.
Thus, 
\begin{equation}
S_{\rm f} (T) \stackrel{T\to \infty}{\longrightarrow}
\prod_{\alpha,n} \left(
2 \Big[
\big( \phi_{n}^{(-)} \big)^2
+
\big( \phi_{-n}^{(-)} \big)^2
\Big]
\right)^{-\left(d_{n}^{\alpha\dagger} d^\alpha_{-n} -\frac{1}{2}\right)}.
\end{equation}

%%%%%%%%%%%%%%%%%%%%%%%%%%%%%%%%%%%%%%%%%%%%%%
\section{S-matrices as two-point functions}
To summarize the result of the previous section, 
$S$-matrix $S(T)=S_{\rm b}(T) S_{\rm f}(T)$ from $\tau=-T$ to $\tau=T$ is 
given by
\begin{eqnarray}
S(T)\! &\!=\!&\! \prod_{I,n}\exp\! \left[-\Omega_{I,n} 
\left(b^\dagger_{I,n}(T) \, b_{I,n}(T)  + \ \frac{1}{2}\right) \right]
\prod_{\alpha,n}\exp\! \left[-\Omega^{\rm f}_{n} (T)
\left(c^{\alpha\dagger}_{n}(T) \, c^{\alpha}_{-n}(T) 
 - \ \frac{1}{2}\right) \right]
\nonumber\\
&\stackrel{T\to \infty}{\longrightarrow} &
\prod_{I,n} \left(
2 f^{(-)}_{I,n}(T) \dot{f}^{(-)}_{I,n}(T) 
\right)^{-\left(a_{I,n}^\dagger a_{I,n} +\frac{1}{2}\right)}
\prod_{\alpha,n} \left(
2 \Big[
\big( \phi_{n}^{(-)} \big)^2
+
\big( \phi_{-n}^{(-)} \big)^2
\Big]
\right)^{-\left(d_{n}^{\alpha\dagger} d^\alpha_{-n} -\frac{1}{2}\right)}
.
\label{eq:Smatsum}
\end{eqnarray}
In \cite{Asano:2003xp,Asano:2004vj},
 we mainly analyze supergravity $n=0$ modes
and find that
\begin{eqnarray}
S^{(n=0)}(T) &=&
\prod_{i=1}^{p+1} e^{-2\left(a_{x_i,n}^\dagger a_{x_i,n} 
+\frac{1}{2}\right)T}
\prod_{l=1}^{7-p} 
e^{- \frac{4}{5-p}\left(a_{y_l,n}^\dagger a_{y_l,n} +\frac{1}{2}\right)T}
\prod_{\alpha=1}^{8} 
e^{- \frac{7-p}{5-p}
\left(d_{0}^{\alpha\dagger} d^\alpha_{0} -\frac{1}{2}\right)T}
\nonumber\\
&=& \exp\left[\left(
-2 N_0^{b,x} -\ft{4}{5-p} N_0^{b,y}
-\ft{7-p}{5-p} N_0^{f} -\ft{(3-p)^2}{5-p}
\right) T
\right]
\end{eqnarray}
for $T\to \infty$
where $N_0^x$ and $N_0^y$ are the total occupation numbers of $n=0$ modes for 
$x_i$ and $y_l$ oscillators respectively. 
Also, $N^{f}_0$ is the number of fermionic oscillators. 
Including the classical and spin angular momentum contribution 
$\exp[ -\frac{4}{5-p} (J+\frac{1}{2}N^{f}_0)T]$ and 
using the relation%
\footnote{ 
We set cutoff at $z(\tau=\pm T) =1/\Lambda$, {\it i.e.,}
$\cosh T = \tilde{\ell} \Lambda$. 
Thus for $T\to \infty$, $e^T \sim 2  \tilde{\ell} \Lambda$.
Recall that $|t_f-t_i|\sim 2\tilde{\ell}$, we see 
$e^T \sim |t_f-t_i| \Lambda$.
We also note that 
we have to keep $\Lambda \sim 1$
since we are in the near-horizon region 
$z \ge 1$ of the background. 
}
 $e^T\sim |t_f-t_i| \Lambda$,
we identify the $S$-matrix with the two-point functions
for a certain operator ${\cal O}$ 
of the boundary gauge theory as
\begin{equation}
\langle \bar{{\cal O}}(t_f){\cal O}(t_i)\rangle
\sim (|t_f-t_i|\Lambda)
^{
-\frac{4}{5-p} (J+\frac{1}{2}N^{f}_0)
-2 N_0^{b,x} -\frac{4}{5-p} N_0^{b,y}
-\frac{7-p}{5-p} N_0^{f} -\frac{(3-p)^2}{5-p}
}.
\label{eq:sugraresult}
\end{equation}
Here ${\cal O}$ is the operator consisting of 
large number $J$ of $Z(=\!\phi_{8-p}\!+\!i\phi_{9-p})$
with 
$N_0^{b,x_i}$, $N_0^{b,y_l}$ and $N^{f,\alpha}_0$ 
 numbers of $D_i$, $\phi_l$'s and $\chi^\alpha$'s
respectively:
They are arranged in a trace symmetrizely.
Here $\phi_i$ $(i=1,\cdots,9-p)$ represent U($N$) adjoint scalars
and $\chi^\alpha$ the half of the 16 spinors of 
the $(p+1)$-dimensional gauge theory.
The result precisely agrees with the analysis from supergravity 
theory \cite{Sekino:1999av,Sekino:2000mg,
Antonuccio:1999iz,Gherghetta:2001iv}.

In the rest of this section, we analyze the 
$n\ne 0$ modes 
using the technique developed in the previous section.

%%%%%%%%%%%%
\subsection{Perturbative analysis for $n\ne 0$ modes }

To obtain $S$-matrix (\ref{eq:Smatsum}) explicitly, 
we need general solutions for the 
equations of motion (\ref{eq:EOMbos}) and (\ref{eq:EOMfer}).
For $n\ne 0$ and $p\ne 3$, 
it is difficult to obtain exact solutions%
\footnote{
For $p=4$, the equation of motion for bosonic fluctuation
(\ref{eq:EOMbos}) can be solved by Mathieu functions.
%~\cite{}.
%To extract the result, however, we need 
%perturbative analysis after all.
}.
To obtain  asymptotic solutions for $\tau\to \infty$
is easier. 
However,
it is not enough by itself since we need 
solutions satisfying the time reflection symmetry.
Here we instead use the perturbative method by using 
$|n|/\tilde{\alpha}\,(=\frac{2}{5-p} \frac{|n|L^2}{J} )$ as an expansion parameter.
%%%
\paragraph{Bosonic case}
%%%
By expanding the field $X_n(\tau)$ with $n\ne 0$ as
\begin{equation}
X_n = X_n^{(0)} 
+ \left(\frac{n}{\tilde{\alpha}}\right)^2 X_n^{(2)}
+\cdots+ 
\left(\frac{n}{\tilde{\alpha}}\right)^{2i} X_n^{(2i)}
+\cdots,
\end{equation}
equations of motion (\ref{eq:EOMbos}) at the order 
$(n/\tilde{\alpha})^{2i}$ for $i\ge 0$ is written as
\begin{equation}
\ddot{X}^{(2i)}_n -\bar{r}(\tau)^{p-3} X^{(2i-2)}_n -m^2 X_n^{(2i)} =0
\end{equation}
where we assign $X^{(-2)}_n =0$.
These are solved recursively as 
\begin{equation}
X_n^{(2i)} =\frac{1}{2 m}
\left[ e^{m \tau} 
\left( \int e^{-m \tau} \bar{r}^{p-3} X_n^{(2i-2)} d\tau
\right)
-  e^{-m \tau} 
\left( \int e^{m \tau} \bar{r}^{p-3} X_n^{(2i-2)} d\tau
\right)
\right].
\end{equation}
In terms of $f_n^{(\pm :2i)}$ with time reflection symmetry 
$f_n^{(+:2i)}(\tau)=f^{(-:2i)}_n (-\tau)$,
solutions are written as
\begin{eqnarray}
f_n^{(\pm :2i)}(\tau) &=& \frac{1}{2 m}
\Bigg[ e^{m \tau} 
\left( \int_0^\tau e^{-m \tau} \bar{r}^{p-3} f_n^{(\pm :2i-2)} d\tau
+ c_n^{(\pm,2i)}
\right)
\nonumber\\
&&\hspace*{3.5cm} +\,  e^{-m \tau} 
\left(- \int_0^\tau e^{m \tau} \bar{r}^{p-3} f_n^{(\pm :2i-2)} d\tau
+ c_n^{(\mp,2i)}
\right)
\Bigg]
\end{eqnarray}
where $c_n^{(\pm,2i)}$ are constants.
These constants can be usually determined by 
the normalization condition (\ref{eq:normcondbos})
and the boundary condition.
%We will analyze the results up to the order $(n/\tilde{\alpha})^2$.
At the 0-th order, $c_n^{(+,0)}=0$ from the boundary condition 
at large $T$. 
Then the normalization condition up to the order 
$(n/\tilde{\alpha})^2$ is 
\begin{eqnarray}
1 &=& f_n^{(+)}  \dot{f}_n^{(-)}  
- f_n^{(-)}  \dot{f}_n^{(+)} 
\nonumber\\
&=& \frac{1}{2m} (c_n^{(-,0)})^2 + 
\frac{1}{m} c_n^{(-,0)} c_n^{(-,2)}
\left(\frac{n}{\tilde{\alpha}}\right)^2 + 
{\cal O} \left( \Big(\frac{n}{\tilde{\alpha}} \Big)^4 \right).
\end{eqnarray}
Thus, we can choose  
\begin{equation}
   c_n^{(-,0)} = \sqrt{2m}, \quad c_n^{(-,2)} =0. 
\end{equation}
For $p<4$, the boundary condition  
$ f_n^{(+ :2)}(\tau \!\to\! \infty) \to 0$ 
is satisfied if we choose 
$c_n^{(+,2)}$ as
\begin{equation}
c_n^{(+,2)} = \frac{1}{\sqrt{2m}}
\int_{\infty}^0 e^{-2m\tau} \bar{r}^{p-3} d\tau .
\end{equation} 
For $p=4$, 
$f_n^{(+ :2)}/f_n^{(- :2)}|_{\tau\to \infty}\to 0$ is satisfied 
for any value of $c_n^{(+,2)}$, 
though $f_n^{(+ :2)}(\tau \!\to\! \infty)$ itself diverges.
In fact, the value of $c_n^{(+,2)}$ does not affect the final result
of $S$-matrix at large $T$.

To summarize, $f_n^{(\pm)}$ up to the order $(n/\tilde{\alpha})^2$ 
for $p< 4$ is 
\begin{equation}
f_{n(p < 4)}^{(\pm)} = \frac{1}{\sqrt{2m}} e^{\mp m \tau} \pm
\frac{1}{(2m)^{3/2}} 
\Bigg[
 e^{\pm m \tau} 
\! \int_{\pm \infty}^\tau \! e^{\mp 2 m \tau} \bar{r}^{p-3}  d\tau
-  e^{\mp m \tau} 
\! \int_{0}^\tau \! \bar{r}^{p-3}  d\tau
\Bigg]
\! \left(\frac{n}{\tilde{\alpha}}\right)^2 +\cdots .
\end{equation}
For $p<3$, 
\begin{equation}
2  f^{(-)}_{n}(T) \dot{f}^{(-)}_{n}(T) 
\stackrel{T\to \infty}{\longrightarrow}
e^{2mT} \left[
1+\frac{1}{4m}(2\ell)^{\frac{2(3-p)}{5-p}}
B\!\left( \ft{3-p}{5-p} ,\ft{3-p}{5-p} \right)
\left(\frac{n}{\tilde{\alpha}}\right)^2
\right] + \cdots 
\end{equation}
and the $S$-matrix up to the order $(n/\tilde{\alpha})^2$ is 
\begin{equation}
S_{n}^{p<3} (T) \sim
\left(
(|t_f-t_i|\Lambda)^{2m}\!
 \left[
1+\ft{1}{4m}
\left(\ft{5-p}{2}\right)^{\frac{2(3-p)}{5-p}}
B\!\left( \ft{3-p}{5-p} ,\ft{3-p}{5-p} \right)
\left(\ft{n}{\tilde{\alpha}}\right)^2
|t_f-t_i|^{\frac{2(3-p)}{5-p}}
\right] 
\right)^{-\left(a_{n}^\dagger a_{n} +\frac{1}{2}\right)}.
\end{equation}
Here $B(p,q)$ is the Beta function defined by
$$
B(p,q) = \int^{\infty}_0 \frac{x^{q-1}}{(1+x)^{p+q}} dx.
$$
In general, $S_n^{p<3}$ is represented as 
\begin{equation}
S_{n}^{p<3} (T) \sim
\left(
(|t_f-t_i|\Lambda)^{2m}\!
 \left[
\sum_{i\ge 0} |t_f-t_i|^{\frac{2(3-p)}{5-p}i}
 \left(\frac{L^4n^2}{J^2}\right)^i
g_i^{(b)} 
\right] 
\right)^{-\left(a_{n}^\dagger a_{n} +\frac{1}{2}\right)}
\label{eq:ple3UVb}
\end{equation}
where $g_i^{(b)}$ are numerical constants
with $g_0^{(b)}=1$ and 
$g_1^{(b)}=\frac{1}{4m}(\frac{5-p}{2})^{-\frac{4}{5-p}}
B(\frac{3-p}{5-p} ,\frac{3-p}{5-p} )$.
This expansion is valid for small
\begin{equation}
 |t_f-t_i|^{\frac{2(3-p)}{5-p}}
 \frac{L^4n^2}{J^2} .
\end{equation}
For comparison, the result for $p=3$ is  
\begin{eqnarray}
2  f^{(-)}_{n}(T) \dot{f}^{(-)}_{n}(T) 
&=& e^{2T} \left(1+ \left(\frac{n}{\tilde{\alpha}}\right)^2 T \right)
+\cdots
\nonumber\\
&\sim& 
(|t_f-t_f|\Lambda)^{2}
\left[
1+ \left(\frac{n}{\tilde{\alpha}}\right)^2 \ln(|t_f-t_f|\Lambda)
\right] +\cdots
\nonumber\\
&\sim& 
%\exp{\left(2 T + \left(\ft{n}{\tilde{\alpha}}\right)^2 T \right)} +\cdots.
(|t_f-t_i|\Lambda)^{2+ \left(\ft{n}{\tilde{\alpha}}\right)^2 +\cdots}.
\end{eqnarray}
This is naturally given by the expansion of 
the known exact result 
$ (|t_f-t_f|\Lambda)^{2 \sqrt{1+(n/\alpha)^2} }$.

For $p=4$, 
\begin{equation}
2  f^{(-)}_{n}(T) \dot{f}^{(-)}_{n}(T) 
= e^{2mT} \left[1+ \ft{1}{m}  \left(\ft{n}{\tilde{\alpha}}\right)^2 
\tilde{\ell}^{-2} (2T +\sinh(2T))
 \right]+\cdots
\end{equation}
and thus 
\begin{equation}
S_{n}^{p=4} (T) \sim
\left(
(|t_f-t_i|\Lambda)^{2m}\!
 \left[
1+ \frac{2}{m}
\Lambda^2 
\left(\frac{L^2 n}{J}\right)^2
+\cdots
\right] 
\right)^{-\left(a_{n}^\dagger a_{n} +\frac{1}{2}\right)}.
\label{eq:p4IRb}
\end{equation}
This expansion is valid if $L^2 |n|/J $ is small since
$\Lambda\sim 1$.
% ($\tilde{\ell}\Lambda \to \infty$).

%interpretation, anomalous dim , )

%%% fermion %%%
%%%
\paragraph{Fermionic case}
%%%
For fermions, we expand the fields $\theta_{\pm,n}$ as 
\begin{eqnarray}
\tilde{\theta}_{+,n} &=& \theta_{+,n}^{(0)} 
+ \frac{n}{\tilde{\alpha}} \, \theta_{+,n}^{(1)}
+\cdots+ 
\left(\frac{n}{\tilde{\alpha}}\right)^{i} \theta_{+,n}^{(i)}
+\cdots, ,
\nonumber\\
{\theta}_{-,n} &=& \theta_{-,n}^{(0)} 
+ \frac{n}{\tilde{\alpha}} \, \theta_{-,n}^{(1)}
+\cdots+ 
\left(\frac{n}{\tilde{\alpha}}\right)^{i} \theta_{-,n}^{(i)}
+\cdots,
\end{eqnarray}
where $ \tilde{\theta}_{+,n}=  i \gamma_{(p)} \theta_{+,n}$.
The equations of motion (\ref{eq:EOMfer}) at the order 
$(n/\tilde{\alpha})^i$ is
\begin{equation}
\dot{\theta}_{\pm,n}^{(i)} \pm \bar{r}^{\frac{p-3}{2}}
{\theta}_{\pm,n}^{(i-1)} -m_{f(p)} {\theta}_{\mp,n}^{(i)}=0  
\end{equation}
with ${\theta}_{\pm,n}^{(i<0)}=0$.
As for the bosonic case, these are solved recursively.
In terms of $ \phi_{n}^{(+:i)}$ and $\psi_{n}^{(+:i)}$ 
of (\ref{eq:fermigensol}),
the solution can be represented as 
\begin{eqnarray}
\phi^{(+:i)}_{n} &=& 
\frac{1}{2}\, e^{m_{f(p)} \tau} \int e^{-m_{f(p)} \tau} 
\bar{r}^{\frac{p-3}{2}} 
\left( \phi_{n}^{(+:i-1)} - {\psi}_{n}^{(+:i-1)} 
\right) d\tau
\nonumber\\
&& \hspace*{1cm}
+
\frac{1}{2} \, e^{- m_{f(p)} \tau} \int e^{m_{f(p)} \tau} 
\bar{r}^{\frac{p-3}{2}} 
\left(  \phi_{n}^{(+:i-1)} + {\psi}_{n}^{(+:i-1)} 
\right) d\tau,
\label{eq:phini}
\end{eqnarray}
\begin{eqnarray}
\psi^{(+:i)}_{n} &=& 
\frac{1}{m_{f(p)}} \left(
\dot{\phi}_{n}^{(+:i)} - \bar{r}^{\frac{p-3}{2}} 
 \phi_{n}^{(+:i-1)}  
\right)
\nonumber\\
&=&
\frac{1}{2} \, e^{m_{f(p)} \tau} \int e^{-m_{f(p)} \tau} 
\bar{r}^{\frac{p-3}{2}} 
\left(  \phi_{n}^{(+:i-1)} - {\psi}_{n}^{(+:i-1)} 
\right) d\tau
\nonumber\\
&& \hspace*{1cm}
-
\frac{1}{2} \, e^{- m_{f(p)} \tau} \int e^{m_{f(p)} \tau} 
\bar{r}^{\frac{p-3}{2}} 
\left(  \phi_{n}^{(+:i-1)} + {\psi}_{n}^{(+:i-1)} 
\right) d\tau . 
\label{eq:psini}
\end{eqnarray}
If we determine $\phi^{(+:i)}_{n\ge 0}$ and 
$\psi^{(+:i)}_{n\ge 0}$ from the above equations, 
the remaining $\phi^{(\pm:i)}_{n<0}$ and $\psi^{(\pm:i)}_{n< 0}$ 
can be obtained by using 
the conditions (\ref{eq:solcond2})
and the time reflection symmetry:
\begin{equation}
\phi_{-n}^{(+:i)} (-\tau) = (-)^i \phi_n^{(-:i)}(\tau);\quad
\phi_n^{(+:i)} =(-)^{i-1} \psi_{-n}^{(+:i)},\quad  
\phi_n^{(-:i)} = (-)^i \psi_{-n}^{(-:i)} .
\end{equation}
In fact, we can adjust the integration constants in 
(\ref{eq:phini}) and (\ref{eq:psini}) 
to satisfy  
\begin{equation}
 \phi_n^{(\pm:i)} =   \phi_{-n}^{(\pm:i)} .
\end{equation}
Then the normalization condition (\ref{eq:solcond1}) reduces to  
\begin{eqnarray}
\frac{1}{2} &=& \phi_n^{(+)} \psi_n^{(-)} - \phi_n^{(-)} \psi_n^{(+)}
\nonumber\\
&=&
2\sum_{i=0}^{\infty} 
\left( \frac{n}{\tilde{\alpha}}\right)^{2i}
\left[ \sum_{j=0}^{2i} \phi_n^{(+:j)} (0) \,
\phi_n^{(+:2i-j)} (0)
\right] .
\end{eqnarray}
By taking into account the boundary condition, 
%$\phi_n^{(+)} \to 0 $
the solution up to the order $(n/\tilde{\alpha})^2$
is
\begin{equation}
\phi_{\pm n}^{(+)}(\tau)
= \phi_{\pm n}^{(+:0)}(\tau)
+ \frac{n}{\tilde{\alpha}} \, \theta_{+,n}^{(1)} 
\phi_{\pm n}^{(+:1)}(\tau)
+ \left(\frac{n}{\tilde{\alpha}}\right)^{2} 
\phi_{\pm n}^{(+:2)}(\tau)+\cdots
\end{equation}
with 
\begin{eqnarray}
\phi_{\pm n}^{(+:0)}(\tau) &=&
 \frac{1}{2} \, e^{  - m_{f(p)} \tau} ,
\\
\phi_{\pm n}^{(+:1)}(\tau) &=&
\frac{1}{2}  \,
e^{ m_{f(p)} \tau} \int_\infty^{\tau} e^{-2 m_{f(p)} \tau} 
\, \bar{r}^{\frac{p-3}{2}} d\tau ,
\\
\phi_{\pm n}^{(+:2)}(\tau) &=&
\frac{1}{2}  \,
e^{- m_{f(p)} \tau} \int_0^{\tau} e^{2 m_{f(p)} \tau} \,
\bar{r}^{\frac{p-3}{2}} 
\left( \int_{\infty}^{\tau} e^{-2 m_{f(p)} \tau} \,
\bar{r}^{\frac{p-3}{2}} d \tau
\right)
d \tau
\nonumber\\
&& \hspace*{10mm} -  \frac{1}{4} \, 
e^{- m_{f(p)} \tau} 
\left( 
\int_0^{\infty} e^{- 2 m_{f(p)} \tau} \,
\bar{r}^{\frac{p-3}{2}} 
d\tau
\right)^2 .
\end{eqnarray}

For $p<3$,
\begin{eqnarray}
2\Big({ \big( \phi_{n}^{(-)}(T) \big)^2
+
\big( \phi_{-n}^{(-)}(T) \big)^2 }\Big)
\nonumber\\
&& \hspace*{-5cm}
\stackrel{T\to \infty}{\longrightarrow}
e^{ 2 m_{f(p)} T} 
\left[
1+ (2 \tilde{\ell})^{\frac{2(3-p)}{5-p}}
\left(\ft{2}{5-p}\right)^{\frac{4}{5-p}} 
\left( N_p - \frac{1}{4} (2^{\frac{4}{5-p}}-1 ) \right)
\left(\frac{n}{\tilde{\alpha}}\right)^{2} 
\right]+\cdots
\end{eqnarray}
where $N_p$ is a constant given by
\begin{equation}
N_p = \frac{2}{5-p} \int^\infty_1 (x^2+1)^{\frac{p-1}{5-p}}
x^{\frac{p-9}{5-p}}.
\end{equation}
Thus the $S$-matrix up to the order $(n/\tilde{\alpha})^2$ is
\begin{eqnarray}
S_{{\rm f};n}^{p<3} (T) &\sim&
\Bigg( (|t_f-t_i|\Lambda)^{\frac{7-p}{5-p} }\!
\nonumber\\ 
&&\hspace*{-9mm}
\times
\left[
1+ |t_f-t_i|^{ \frac{2(3-p)}{5-p} }
\left(\ft{2}{5-p}\right)^{\frac{4}{5-p}} \!
\left( N_p - \frac{1}{4} (2^{\frac{4}{5-p}}-1 ) \right)\!
\left(\frac{n}{\tilde{\alpha}}\right)^{2} 
\right]\Bigg)^{-\left(d_n^\dagger d_{-n}-\frac{1}{2} \right)}
\!\!\!\!.
\end{eqnarray}
As for the bosonic case, 
if $ |t_f-t_i|^{\frac{2(3-p)}{5-p}}
 \frac{L^4n^2}{J^2}$ is small enough, we can in principle 
represent $ S_{{\rm f};n}^{p<3} (T)$ as
\begin{equation}
 S_{{\rm f};n}^{p<3} (T) \sim
\left(
(|t_f-t_i|\Lambda)^{\frac{(7-p)}{5-p}   }\!
 \left[
\sum_{i\ge 0} |t_f-t_i|^{\frac{2(3-p)}{5-p}i}
 \left(\frac{L^4n^2}{J^2}\right)^i
g_i^{(f)} 
\right] 
\right)^{-\left(d_{n}^\dagger d_{-n} +\frac{1}{2}\right)}
\label{eq:ple3UVf}
\end{equation}

For $p=4$,
\begin{equation}
2\Big[{ \big( \phi_{n}^{(-)}(T) \big)^2
+
\big( \phi_{-n}^{(-)}(T) \big)^2 }\Big]
\stackrel{T\to \infty}{\longrightarrow}
e^{3T} \left[1+ \frac{5}{16} \tilde{\ell}^{-2}
e^{2T}\left(\frac{n}{\tilde{\alpha}}\right)^{2}
\right] +\cdots 
\label{eq:nexp4}
\end{equation}
and we see that if $L^2 |n|/J$ is small, 
the $S$-matrix for this part becomes 
\begin{equation}
S_{{\rm f};n}^{p=4} (T) \sim
\left( (|t_f-t_i|\Lambda)^3 \left[1+ \frac{5}{16}
% |t_f-t_i|^{-2} (|t_f-t_i|\Lambda)^2
\Lambda^2
\left(\frac{n}{\tilde{\alpha}}\right)^{2}
+\cdots
\right]\right)^{-\left(d_n^\dagger d_{-n}-\frac{1}{2} \right)}.
\label{eq:p4IRf}
\end{equation}

For $p=3$, we obtain the expansion of the known exact result 
\begin{equation}
S_{{\rm f};n}^{p=3} (T) \sim
(|t_f-t_i|\Lambda)^{-\sqrt{1+(n/\alpha)^2} (2 d_n^\dagger d_{-n} -1) }
\end{equation}
for the same analysis.

%%%%%%%%%%%%
\subsection{$|n|/\tilde{\alpha}\to \infty$ limit}
We will briefly give the analysis of $S$-matrix at 
large $(n/\tilde{\alpha})^2$.
For this purpose, it is convenient to rewrite the equations of 
motion 
% (\ref{eq:EOMbos}) and (\ref{eq:EOMfer})
by making the redefinition of fields 
$(x_i,y_l)\to (X_i,Y_l)$ and $(\tau,\sigma)\to (\tau_c,\sigma_c)$
by
\begin{equation}
X_i = \ft{2}{5-p} \bar{r}^{-{3-p\over 4}} x_i,
\quad
Y_l = \bar{r}^{-{3-p\over 4}} y_l,
\end{equation}
\begin{equation}
\sigma = \ft{5-p}{2}\ell \sigma_c \,, \quad 
{d\tau \over d\tau_c} = \ft{5-p}{2}\ell \bar{r}^{3-p\over 2}\,. 
%\label{eq:wsredef}
\end{equation}
Note that this definition of the world-sheet fields 
corresponds to the one 
with the conformal gauge fixing 
$\sqrt{h}h^{\alpha\beta}=\delta^{\alpha\beta}$.
Then the equations motion for bosonic fields (\ref{eq:EOMbos}) 
become 
\begin{equation}
\frac{d^2 {X}_n}{d\tau_c^2} 
- \left[ \left(\frac{n}{\alpha}\right)^2 \ell^2 +m_X(\tau_c)^2  
 \right]
X_n =0
\end{equation}
where $\alpha=J/L^2$ and
\begin{eqnarray}
m_X^{2}(\tau_c) &=& -
{(7-p) \over 16 \bar{r}^{2}}
[ (3-p) +(3p-13)\ell^{2} \bar{r}^{5-p}  ]
,
%\label{eq:timedepMX}
\\
m_Y^{2}(\tau_c) &=& - {(7-p) \over 16 \bar{r}^{2}}
[ (3-p) -(p+1)\ell^{2} \bar{r}^{5-p}  ].
%\label{eq:timedepMY}
\end{eqnarray}
Also, (\ref{eq:EOMfer}) becomes
\begin{equation}
\frac{d}{d\tau_c} {\theta}_{\pm,n} 
\pm \ell \frac{n}{\alpha} {\theta}_{\pm,n} 
-i m_{f(p)} \ft{5-p}{2} \ell \bar{r}^{\frac{3-p}{2}}
\gamma_{(p)} \theta_{\mp,n}
 =  0 .
\end{equation}
For $(n/\alpha)^2 \gg 1$,
these equations of motion 
reduce to 
\begin{equation}
\frac{d^2 {X}_n}{d\tau_c^2} 
- \left(\ell \frac{n}{\alpha}\right)^2  
X_n =0
\qquad
\mbox{and}
\qquad
\frac{d}{d\tau_c} {\theta}_{\pm,n} 
\pm \ell \frac{n}{\alpha} {\theta}_{\pm,n} 
 =  0 ,
\end{equation}
which are immediately solved as
\begin{equation}
X_n = C_{\pm} \exp \left( \pm \frac{n}{\alpha} \ell  \tau_c \right)
\qquad
\mbox{and}
\qquad
\theta_{\pm,n}^{\alpha} = \tilde{C}_{\pm}^{\alpha}
 \exp \left( \mp \frac{n}{\alpha} \ell  \tau_c \right).
\end{equation}
Thus, in terms of $f_{n}^{(\pm)}$ and 
$\phi_{n}^{(\pm)}$,
\begin{equation}
f^{(\pm)}_{n}= \sqrt{\frac{\alpha}{2|n|\ell}}
   \exp \left( \mp \frac{|n|}{\alpha} \ell  \tau_c \right)
\quad
\mbox{and}
\quad
(\phi_{n}^{(+)}, \phi_{n}^{(-)})=
\left(0, -\frac{1}{\sqrt{2}}\exp 
\left( \frac{|n|}{\alpha} \ell  \tau_c \right)
 \right).
\end{equation}
Then the $S$-matrix for bosonic part becomes
\begin{eqnarray}
S^{p<3}_{{\rm b}, n}(T_c) 
 &\stackrel{T_c \to T_c^{(\infty)}\, (T\to \infty)}{\longrightarrow}&
\exp \left[ 2\ell \frac{|n|}{\alpha} T_c^{(\infty)}
\left(a_n^\dagger a_n +\ft{1}{2}\right)
 \right]
\nonumber\\
&\sim& \hspace*{-7mm}
\exp\left[
-2 {c^{(0)}} \frac{L^2|n|}{J} 
|t_f-t_i|^{\frac{3-p}{5-p}} \left(a_n^\dagger a_n +\ft{1}{2}\right)
\right]
\label{eq:ple3IRb}
\end{eqnarray}
where $ c^{(0)}\, (<\infty) $  is a constant given by
\begin{eqnarray}
 T_c^{(\infty)} &=&
\int^{\infty}_{\ell^{-2/(5-p)}} \frac{1}{\sqrt{\ell^2 r^{5-p} -1}}dr
\nonumber\\
&=&
\ell^{-\frac{2}{5-p}} \frac{1}{5-p}
\int^{\infty}_1
\frac{1}{\sqrt{u-1}} u^{-\frac{4-p}{5-p}}du 
\nonumber\\
&\equiv & \ell^{\frac{2}{5-p}}\,  c^{(0)} .
\label{eq:defc0nl}
\end{eqnarray} 
For $p=4$, 
\begin{eqnarray}
S^{p=4}_{{\rm b}, n}(T_c) 
 &\stackrel{T_c \to T_c^{(\infty)}}{\longrightarrow}&
\exp \left[- 2\ell \frac{|n|}{\alpha} T_c^{(\infty)}
\left(a_n^\dagger a_n +\ft{1}{2}\right)
 \right]
\nonumber\\
&\sim& \hspace*{-7mm}
\exp\left[
-8 \frac{L^2|n|}{J} 
\Lambda
\left(a_n^\dagger a_n +\ft{1}{2}\right)
\right]
\end{eqnarray}
where we have used  
$\ell T_c^{\infty}  \sim 4 \Lambda$ at 
$z= 1/\Lambda$ ($T\to \infty$),
which comes from the relation 
$\bar{r}(\tau_c)= \frac{\ell^2}{4}\tau_c^2 +\frac{1}{{\ell^2}}$ 
for $p=4$. 
We see that there is no $|t_f-t_i|$ dependence in $S^{p=4}_{{\rm b}, n}$.

We can calculate the contribution of $ \alpha/|n|$ 
by using the similar perturbative method given in the previous
subsection.
The result for $p=4$ is 
\begin{equation}
S^{p=4}_{{\rm b},n} \sim 
\exp\left\{
\left[
-8 \frac{L^2|n|}{J} 
\Lambda
- \frac{3}{128} c^{(1)} \frac{J}{L^2|n|} |t_f-t_i|
+\cdots \right]
\left(a_n^\dagger a_n +\ft{1}{2}\right)
\right\}
\label{eq:p4UVb}
\end{equation}
where $c^{(1)}=1$ for $x_i$ and $c^{(1)}=9$ for $y_l$.
On the other hand, for $p<3$, the $ \alpha/|n|$ correction 
gives only a contribution without $|t_f-t_i|$ dependence,
which is the same phenomena as the cases of $|n|/\alpha$
expansion for $p=4$ given in (\ref{eq:nexp4}).

Similarly, $S_{{\rm f},n}$ for fermionic part becomes
\begin{equation}
S^{p<3}_{{\rm f}, n}(T_c) 
\stackrel{T_c \to T_c^{(\infty)}}{\longrightarrow}
\exp\left[
-2 {c^{(0)}} \frac{L^2|n|}{J} 
|t_f-t_i|^{\frac{3-p}{5-p}} \left(b_n^\dagger b_{-n} -\ft{1}{2}\right)
\right]
\label{eq:pl3flead}
\end{equation}
and
\begin{equation}
S^{p=4}_{{\rm f},n} 
\stackrel{T_c \to T_c^{(\infty)}}{\longrightarrow}
\exp
%\left\{
\left[
-8 \frac{L^2|n|}{J} 
\Lambda
%- \frac{3}{128} c_f \frac{J}{L^2|n|} |t_f-t_i|
%+\cdots 
\right]
\left(b_n^\dagger b_{-n} - \ft{1}{2}\right)
%\right\}
.
\label{eq:p4UVf}
\end{equation}
Note that the constant $c^{(0)}$ appearing in (\ref{eq:pl3flead})
is the same as for the bosonic part given in (\ref{eq:defc0nl}).

We can also compute the $\alpha/|n|$ correction to 
the above result: 
The correction terms for each $p$
has the same form as the bosonic case,
though the coefficients corresponding to $c^{(1)}$ are different 
from bosonic and fermionic contributions.

%%%%%
\subsection{Interpretation as gauge theory correlators}
We briefly discuss the interpretation of  
the above results as 
two-point functions between $(t_f,x^a_f)=(t_f,0) $
and $(t_i,x^a_i)=(t_i,0) $
for the dual gauge theory.

First, for $p\ne 3$, we see that 
there appears a dimensionful quantity
$
n^2/\alpha^2 
%(=(\ft{5-p}{2})^2 n^2/\tilde{\alpha}^2 ) 
= L^4 n^2/J^2
$
(or 
${n^2}/{\tilde{\alpha}^2}= 
(\frac{2}{5-p})^2 {n^2 L^4}/{J^2} 
$) 
in the $S$-matrix 
other than $|t_f-t_i|=2\tilde{\ell}$.
Thus, we characterize the IR or UV behavior of the 
dual gauge theory by measuring the dimensionless 
quantity
\begin{equation}
|t_f-t_i|^{\frac{2(3-p)}{5-p}} \frac{L^4 n^2}{J^2}.
\end{equation}

For $p<3$, if $|t_f-t_i|^{\frac{2(3-p)}{5-p}} \frac{L^4 n^2}{J^2}$
is small, {\it i.e.,} UV, then the expansion with respect to 
$|n|/\alpha$ is valid since we consider the situation 
$ |t_f-t_i|\Lambda \to \infty$ and $\Lambda\sim 1$.
In this case, the corresponding two-point functions are 
given by (\ref{eq:ple3UVb}) and (\ref{eq:ple3UVf})
with classical and spin angular momentum contribution 
$ (|t_f-t_i|\Lambda)^{ -\frac{4}{5-p} (J+\frac{1}{2}N^{\rm f})}$.
On the other hand, the result for $|n|/\alpha\to \infty$ 
represents the IR property%
\footnote{
In the IR limit for $p<3$, 
the (tunneling) null geodesic approaches $r\to 0$
where the dilaton expectation value diverges
and we cannot neglect the string loop effect there.  
This corresponds to the $1\over N$-correction
in the $N\to \infty$ limit with large $g_s N$. 
}.

For $p=4$, 
the expansion with respect to small 
$|n|/\alpha$ represents the IR behavior:
The resulting $S$-matrix is given by
 (\ref{eq:p4IRb}) and (\ref{eq:p4IRf}) and we see that 
the (normalized) two-point functions are the 
same as for the
supergravity modes $n=0$.
If $|t_f-t_i|^{\frac{2(3-p)}{5-p}} \frac{L^4 n^2}{J^2}$
is large, {\it i.e.,} UV, 
then $|n|/\alpha \to \infty$. 
In this case, the corresponding two-point functions 
behave like massive particles as we see from 
(\ref{eq:p4UVb}).
% and (\ref{eq:p4UVf}) 
However, the result does not mean the
appearance of finite correlation length
since the equation does not hold for long distance.

\medskip

For $p=3$, the expansion of two-point functions 
with respect to $L^4  n^2 /J^2$ 
corresponds to the perturbative expansion of the gauge theory side
since $L^4 \sim g_{\rm YM}^2 N $: 
At the $n^2$ order,  
this effect appears as 
$\sim n^2g_{\rm YM}^2 N /J^2 \ln (|t_f-t_i|\Lambda) $
which corresponds to the 
explicit perturbative calculation of 
the gauge theory~\cite{Berenstein:2002jq, Santambrogio:2002sb}. 
On the other hand, for $p\ne 3$, 
the expansion with respect to $|n|/\alpha$ 
does not correspond to the effect of 
perturbative expansion since 
the combination $|t_f-t_i|^{\frac{2(3-p)}{5-p}} \frac{L^4 n^2}{J^2}$
is written by the original coordinates as
\begin{equation}
(|t_f-t_i|_{\rm orig.})^{ \frac{2(3-p)}{5-p} } 
 (g_{\rm YM}^2 N)^{\frac{2}{5-p} }  
\frac{n^2}{J^2}
\end{equation}
and it has fractional power of the coupling constant.

In practice, it is difficult to check the results from 
the gauge theory side since there are severe 
infra-red ($p<3$) or 
ultra-violet ($p=4$) divergence.
Furthermore, the result
for supergravity modes already has the 
non-trivial form (\ref{eq:sugraresult}): 
We see that the dimension of the 
scalar $\phi_l$ does not correspond to that of free-field. 
For $p=4$, the result shows that 
in the infra-red limit 
the two-point functions for 
any $n$ degenerate to the one for $n=0$ modes.
This means that we have non-trivial infra-red 
fixed points for the corresponding $d=5$ theory 
where free-fields of effective dimension 
$d_{\rm eff}=6$ appear.

\medskip

Also, there is a problem which operator 
we should identify for each sector of $S$-matrix  
including $n\ne 0$ modes.
We assume that we can obtain the {\it diagonalized} two-point functions
from the string $S$-matrix by our method.
For $p=3$, we know that the BMN-operators such as
$$
{\cal O} \sim \sum_{j=1}^J e^{\frac{2\pi i j n}{J}}
{\rm Tr} [Z^j \phi_a Z^{J-j}\phi_b ]
$$
give the correct result.
For $p\ne 3$, we may expect that the similar 
BMN-type operators 
play the role as the operators with diagonalized 
two-point functions. 
(For supergravity modes $n=0$, this choice is consistent 
with the analysis from supergravity theories.)
However, as we have stated above, 
it is difficult to check them 
%without powerful tools 
in the gauge theory side. 

\medskip

Finally, we comment on the effect of zero-point energies
from stringy modes.
In the analysis of supergravity modes, 
the results including the effect of zero-point energies 
agree with the analysis 
from supergravity theory~\cite{Asano:2004vj}.
If the result is true, 
the zero-point energies from $n\ne 0$ modes must cancel
for bosonic and fermionic fluctuations.
For $p<3$, the effects of zero-point energies 
from bosonic and fermionic fluctuations 
for small $|n|/\tilde{\alpha}$
can be read from 
(\ref{eq:ple3UVb}) and (\ref{eq:ple3UVf})
respectively.
If this expansion is valid for large $|n|$, the 
corresponding zero-point energies seem to diverge. 
However, for $|n|\to \infty$, the zero-point energies 
vanish as we see from (\ref{eq:ple3IRb}) and (\ref{eq:pl3flead}).
Thus we may expect that the total zero-point energies for 
all $n\ne 0$ modes vanish and that our result is consistent with 
the analysis from the supergravity theory.
The situation for $p=4$ is similar.

%%%%%%%%%%%%%%%%%%%%%%%%%%%%%%%%%%%%%%%%%%%%%%
\section{Concluding remarks}
We have investigated the $S$-matrix for superstring fluctuations around the 
null geodesic of the D$p$-brane background ($0\le p\le 4$)
and give a prediction to the 
two-point functions for 
BMN type operators in the $(p+1)$-dimensional gauge theory
by assuming the holographic correspondence.
In particular, we have studied the effect of string higher modes 
$n\ne 0$ that had remained to be analyzed in the previous 
papers~\cite{Asano:2003xp,Asano:2004vj}.
The main results for two-point functions 
are collected in section 4.3.

%zero-point energy for $n\ne 0$ modes 

With the analysis given above, 
we have almost completed the analysis 
of two-point functions from the string theory side.
The remaining task is,
as we have emphasized repeatedly in our previous papers, 
 to analyze the correlation functions 
in terms of the dual gauge theory itself,
though it would be a non-trivial task to perform.

Finally, since we have a definite procedure to obtain the 
two-point functions from the 
string $S$-matrix around geodesic 
respecting the holographic principle, 
it is easy to apply our method to other cases.
In particular, it would be interesting to 
consider the non-BPS 
expanding strings like 
spinning strings~\cite{Frolov:2002av, Frolov:2003tu}.

%%%%%%%%%%%%%%%%%%%%%%%%%%%%%%%%%%%%%%%%%%%%%%
\vspace{0.2cm}
\paragraph{Acknowledgements}
We would like to thank Y. Sekino and T. Yoneya 
for various discussions and comments. 
We also would like to thank M.~Fukuma, H.~Hata and H.~Kawai
for useful discussions.
This work is supported in part by the Grants-in-Aid for 
the 21st Century COE ``Center for Diversity and Universality in Physics'' 
from the Ministry of Education, Culture, Sports, 
Science and Technology (MEXT) of Japan.

%%%%%%%%%%%%%%%%%%%%%%%%%%%%%%%%%%%%%%%%%%%%%%
%\section*{Appendix}

\appendix 
%%%%%%%%%%
\section{Diagonalization of $S$-matrix from fermionic oscillators}
\setcounter{equation}{0}
\renewcommand{\theequation}{\Alph{section}.\arabic{equation}}
%%%%%%
We will briefly explain the relation 
between two expressions (\ref{eq:normalof}) and
(\ref{eq:expof}) 
of the fermionic part of $S$-matrix $S_{\rm f}$ 
and perform `diagonalization' by $T$-dependent 
 Bogoliubov transformation.

%%%
\paragraph{Normal-ordered form $\to$ Exponential form}
%%%
In order to see the relation between 
$(A^{\rm f}_n, B^{\rm f}_n )$ in (\ref{eq:normalof}) and
$(\tilde{A}^{\rm f}_n, \tilde{B}^{\rm f}_n )$ in 
(\ref{eq:expof}),
we calculate 
$$
{V_n} \equiv \left(
\begin{array}{c}
S_{\rm f} d_n S_{\rm f}^{-1}
\\ 
S_{\rm f} d_n^\dagger S_{\rm f}^{-1}
\end{array}
\right)
$$ 
in both representations.
The results are respectively written as 
\begin{equation}
V_n = \frac{1}{1+B_{n}}
\left(
\begin{array}{cc}
1 & A_{n}
\\ 
A_{n} & (1+B_{n})^2 + A_{n}^2 
\end{array}
\right)
\left(
\begin{array}{c}
d_n 
\\ 
d_n^\dagger 
\end{array}
\right)
\end{equation}
and 
\begin{eqnarray}
V_n &=& 
\exp
\left(
\begin{array}{cc}
-\tilde{B}_{n} & \tilde{A}_{n}
\\ 
\tilde{A}_{n} & \tilde{B}_{n} 
\end{array}
\right)
\left(
\begin{array}{c}
d_n 
\\ 
d_n^\dagger 
\end{array}
\right)
\nonumber\\
&=&
\left[
\cosh\Omega_n + \left(
\begin{array}{cc}
-\tilde{B}_{n} & \tilde{A}_{n}
\\ 
\tilde{A}_{n} & \tilde{B}_{n} 
\end{array}
\right)
\frac{\sinh\Omega_n}{\Omega_n}
\right]
\left(
\begin{array}{c}
d_n 
\\ 
d_n^\dagger 
\end{array}
\right)
.
\label{eq:Vnexp}
\end{eqnarray}

To relate $N_n$ and $\tilde{N}_n$,
we calculate $\langle 0 | S_{\rm f} |0 \rangle$ in both
representations and as a result we have 
\begin{equation}
\prod_{\alpha,n} \! N^{\rm f}_n
=
\left(
\prod_{\alpha, n} \! \tilde{N}^{\rm f}_n
\right)
\langle 0 |
\sum_{\alpha,n}
\left[{1\over 2} \tilde{A}^{\rm f}_{n} 
d^{\alpha\dagger}_{n} d^{\alpha\dagger}_{-n} 
+ \tilde{B}^{\rm f}_{n} d^{\alpha\dagger}_{n} d^{\alpha}_{-n}  
+ {1\over 2} \tilde{A}^{\rm f}_{n} d^{\alpha}_{n} d^{\alpha}_{-n} 
\right]
 |0 \rangle
\end{equation}
This equation can be simplified 
by using the following technique.
We define 
\begin{equation}
 S_{\rm f}(\epsilon) \equiv 
\left(\prod_{\alpha,n} \tilde{N}^{\rm f}_{n}\right)
 \exp\left\{\epsilon
\sum_{\alpha,n}
\left[{1\over 2} \tilde{A}^{\rm f}_{n}
d^{\alpha\dagger}_{n} d^{\alpha\dagger}_{-n} 
+ \tilde{B}^{\rm f}_{n} d^{\alpha\dagger}_{n} d^{\alpha}_{-n}  
+ {1\over 2} \tilde{A}^{\rm f}_{n} d^{\alpha}_{n} d^{\alpha}_{-n} 
\right]\right\} 
\end{equation}
and differentiate $ \langle 0 | S_{\rm f}(\epsilon) |0 \rangle $ 
with respect to $\epsilon$:
\begin{equation}
\frac{d}{d\epsilon}
\langle 0 | S_{\rm f}(\epsilon) |0 \rangle 
=
\left(
\prod_{\alpha,n} \! \tilde{N}^{\rm f}_n
\right)
\langle 0 | S_{\rm f}(\epsilon) 
\sum_{\alpha,n}\ft{1}{2}
\tilde{A}^{\rm f}_{n}
d^{\alpha\dagger}_{n} d^{\alpha\dagger}_{-n} 
|0 \rangle .
\label{eq:diffSfe}
\end{equation}
By using the relation corresponding to (\ref{eq:Vnexp}) with 
$\epsilon$,
the right-hand side is rewritten as the form 
$
\mbox{[coefficients]} \times \langle 0 | S_{\rm f}(\epsilon) |0 \rangle 
$
and we can solve (\ref{eq:diffSfe}) as a differential 
equation of $\epsilon$.
The result is 
\begin{equation}
\langle 0 | S_{\rm f}(\epsilon=1) |0 \rangle 
= \prod_{\alpha,n} 
\left( 
\tilde{N}^{\rm f}_n  \exp\left({\tilde{B}^{\rm f}_{n}\over 2}\right) 
\frac{1}{\sqrt{1+B^{\rm f}_{n}}} 
\right).
\end{equation}
In fact we can show that 
\begin{equation}
\tilde{N}^{\rm f}_n =
N^{\rm f}_n  \exp\left(-{\tilde{B}^{\rm f}_{n}\over 2}\right) \sqrt{1+B^{\rm f}_{n}} .
\end{equation}
%
%
%%%
\paragraph{Diagonalization}
We can diagonalize $S_{\rm f}$ in the exponential representation 
as
\begin{eqnarray}
 S_{\rm f} &=&
S_{{\rm f},n=0} \prod_{\alpha, n>0}
(\tilde{N}^{\rm f}_n)^2
\exp
\left[ \tilde{A}^{\rm f}_{n}
d^{\alpha\dagger}_{n} d^{\alpha\dagger}_{-n} 
+ \tilde{B}^{\rm f}_{n} 
( d^{\alpha\dagger}_{n} d^{\alpha}_{-n}  
+d^{\alpha\dagger}_{-n} d^{\alpha}_{n}  
)
+  \tilde{A}^{\rm f}_{n} d^{\alpha}_{n} d^{\alpha}_{-n} 
\right] 
\nonumber\\
&=&
\!
S_{{\rm f},n=0} \! \prod_{\alpha,n>0} \!
(\tilde{N}^{\rm f}_n)^2
\exp(\tilde{B}^{\rm f}_{n} + \Omega_n^{\rm f} )
\exp\!
\left[
- \Omega_n^{\rm f} 
\left(c^{\alpha\dagger}_{n}(T) c^{\alpha}_{-n}(T)  
+c^{\alpha\dagger}_{-n}(T) c^{\alpha}_{n}(T)   
\right)
\right].
\end{eqnarray}
We use $T$-dependent operators 
$(c^{\alpha\dagger}_{n}(T), c^{\alpha}_{-n}(T) )  $
satisfying 
(\ref{eq:anticomc})
defined by 
\begin{equation} 
\left(
\begin{array}{c}
c_{n}^\dagger(T) \\ c_{n} (T)
\end{array}
\right)
=  
\left(
\begin{array}{cc}
G_{n}^{\rm f}(T)  & F_{n}^{\rm f} (T)
\\ E_{n}^{\rm f}(T) & D_{n}^{\rm f}(T) 
\end{array}
\right)
\left(
\begin{array}{c}
d^\dagger_{n} \\ d_{n} 
\end{array}
\right).
\end{equation}
Here $D_{n}^{\rm f}$, $G_{n}^{\rm f}$, $F_{n}^{\rm f}$ and $E_{n}^{\rm f}$
are determined in order to satisfy the relations
$ D_{n}^{\rm f} G_{n}^{\rm f}- F_{n}^{\rm f} E_{n}^{\rm f}  =1$,
$ (D_{n}^{\rm f}, G_{n}^{\rm f}, F_{n}^{\rm f}, E_{n}^{\rm f}  )
=  (D_{-n}^{\rm f}, G_{-n}^{\rm f}, -F_{-n}^{\rm f}, -E_{-n}^{\rm f}  )
 $
 and  
\begin{equation} 
\left(
\begin{array}{cc}
D_{n}^{\rm f}  & \mp E_{n}^{\rm f} 
\\ \mp F_{n}^{\rm f} & G_{n}^{\rm f} 
\end{array}
\right)
\left(
\begin{array}{cc}
\pm \tilde{A}_{n}^{\rm f}  & \tilde{B}_{n}^{\rm f}  
\\ - \tilde{B}_{n}^{\rm f}   & \pm \tilde{A}_{n}^{\rm f}   
\end{array}
\right)
\left(
\begin{array}{cc}
D_{n}^{\rm f}  & \pm F_{n}^{\rm f} 
\\ \pm E_{n}^{\rm f} & G_{n}^{\rm f} 
\end{array}
\right)
=
\left(
\begin{array}{cc}
0  & -\Omega^{\rm f}_n
\\ \Omega^{\rm f}_n & 0
\end{array}
\right).
\end{equation}
If necessary, 
we can diagonalize $S_{{\rm f},n=0}$ separately, though 
it is already diagonalized in our case. 
Thus, the final form of diagonalized $S$-matrix becomes
(\ref{eq:diagSf}).

%%%%%%%%%%%%%%%%%%%%%%%%%%%%%%%%%%%%%%%%%%%%%%

\end{document}